\documentclass[12pt]{article}

\usepackage{epsf,cite,a4wide,latexsym}
\newcommand{\beq}{\begin{equation}}
\newcommand{\eeq}{\end{equation}}
\newcommand{\bea}{\begin{eqnarray}}
\newcommand{\eea}{\end{eqnarray}}
\newcommand{\bda}{\begin{eqnarray*}}
\newcommand{\eda}{\end{eqnarray*}}

\newcommand{\m}{\mu}

\newcommand{\oh}{\frac{1}{2}}

\newcommand{\dg}{\dagger}

\newcommand{\Id}{1\!\!1}
\newcommand{\Tr}{{\rm Tr}}
\newcommand{\Epsilon}{{\cal E}}

\begin{document}

\vskip -4cm

\begin{flushright}
FTUAM-99-16
\end{flushright}

\vskip 0.2cm

{\Large
\centerline{{\bf Study of SU(3) vortex-like configurations}}
\centerline{{\bf with a new maximal center gauge fixing method.}}
\vskip 0.3cm

\centerline{ A. Montero }}
\vskip 0.3cm

\centerline{Departamento de F\'{\i}sica Te\'orica C-XI,}
\centerline{Universidad Aut\'onoma de Madrid,}
\centerline{Cantoblanco, Madrid 28049, SPAIN.}

\vskip 0.8cm

\begin{center}
{\bf ABSTRACT}
\end{center}
We present a new way of fixing the gauge to (direct) maximal
center gauge in SU(N) Yang-Mills theory and apply this
method to SU(3) configurations which are vortex-like.
We study the structure of the $Z_3$ configurations obtained
after center-projecting the SU(3) ones.

\vskip 1.5 cm
\begin{flushleft}

\end{flushleft}

\newpage

\section{Introduction.}

In this paper we present a new algorithm to fix the gauge to direct maximal 
center gauge in lattice SU(N) Yang-Mills theory, and we show how this gauge
fixing works in SU(3) configurations which are vortex-like. 

A number of recent works try to explain low energy phenomena, such
as confinement or chiral symmetry breaking, in terms of vortices, giving
a renewed interest to the vortex condensation theory of confinement proposed
by many authors at the end of 70's \cite{thooft}. In 
\cite{green1,green2,green3} the maximal center gauge is used; this gauge 
selection has the property of making the link variables as close as possible
to the elements of the center of the group. Afterwards the configuration is center-projected
to unravel its (vortex) structure. They conclude that
with  this method it is possible to identify unprojected center vortices, and 
that their results support the picture of these objects being responsible
for quark confinement. Two different methods are used to transform lattice 
configurations to the maximal center gauge. The first one (indirect maximal
center gauge) is a two step procedure: first the link variables are transformed
to the maximal abelian gauge and projected over abelian link variables; and then
the abelian link variables are gauge transformed to new variables as close as
possible to the elements of the group center, and projected over center link variables.
The second one (direct maximal center gauge) tries to obtain the link variables
as close as possible to the elements of the group center in a direct way, without any
intermediate step. In references
\cite{green1,green2,green3}, both methods are used and the conclusion that center vortices
are the relevant objects for confinement is claimed to be
independent of the method. In references \cite{lang,lang2}, following
the same procedures, the vortex content of the SU(2) vacuum is studied
and also the relevance of center vortices at finite temperature. Their conclusion
is that the dynamics of these objects reflects the transition to the
deconfining phase and also that they are the relevant degrees of freedom to determine
the spatial string tension at finite temperature.      
In references \cite{cher1,cher2} the maximal center gauge
is also used, in this case only from the indirect method,
to study confinement in SU(2) and SU(3) Yang-Mills theory. In these papers
a new topological object is constructed, the center monopole, and
the conclusion of these works \cite{cher1,cher2}
is that this object condenses in the confinement phase and
does not condense in the deconfinement phase, its condensate being
therefore the order parameter of the theory.
In reference \cite{forcrand} the relevance of center vortices for confinement and 
chiral symmetry breaking is studied. The result presented in that work is that,
after removing center vortices in an ensemble of SU(2) lattice configurations, confinement is
lost and chiral symmetry is restored. The conclusion is that center
vortices are responsible for both phenomena. In this case the method used to gauge-fix
the configurations is the direct maximal center gauge.
Finally, we  want to mention the works in reference \cite{kovacs} relating center vortices with
confinement but without using the maximal center gauge.

In this work we present an algorithm of direct center gauge fixing for SU(N) lattice 
gauge theory. Our motivation is twofold. First, there is no efficient
method of direct gauge fixing to maximal center gauge in SU(N) lattice gauge theory;
secondly, it is very interesting to know how SU(N) configurations which are vortex-like
look like in the maximal center gauge, and therefore, whether these solutions
share the properties described in 
\cite{green1,green2,green3,lang,lang2,cher1,cher2,forcrand} for a confining object.

The paper is structured as follows: In section 2 the numerical algorithm for
gauge fixing is described. Section 3 shows how this method works on classical solutions
of the SU(3) Yang-Mills theory which are vortex-like, and how these configurations
appear when center-projected. Our conclusions are presented in section 4.

\section{Gauge fixing method.}

   The maximal center gauge in SU(N) lattice gauge theory is defined as the
gauge which brings link variables $U$ as close as possible to elements of its center
$Z_N = \left\{e^{2\pi m i/N}\Id \right.$ , $\left. m=0,...,N-1 \right\}$.
This can be achieved by maximizing one of the following quantities:
\bea
& & R=\frac{1}{N_{site}N_{dim}N^N}\sum_{n,\mu} \mbox{Re} \left( [\mbox{Tr}\; U_\mu(n) ]^N \right)
      \hspace{0.5 cm}, \label{barionlike}  \\
& & R=\frac{1}{N_{site}N_{dim}N^2}\sum_{n,\mu}\mid\mbox{Tr}\;U_\mu(n)\mid^2 
      \hspace{0.5 cm}, \label{mesonlike}
\eea
where $N_{site}$ is the number of sites on the lattice and $N_{dim}$ the number of 
dimensions ($R$ satisfies $|R|\le 1$). In reference \cite{green3}
the quantity given in equation (\ref{barionlike}) is maximized using the method of simulated
annealing but, as it is stated there, this is an inefficient method.
Our choice is that of equation (\ref{mesonlike}) because an update of the link variables
as in the Cabibbo-Marinari-Okawa method \cite{cabibbo,okawa} drives to a simple way to maximize $R$.
This procedure is inspired in the one described in \cite{green3} for SU(2). It is also 
interesting to note that maximizing equation \ref{mesonlike} is equivalent to maximize the
trace in the adjoint representation which is not affected by center gauge transformations. 

We perform a local update of the link variables.
At each site $n$, one needs to find the gauge transformation $G$ which maximizes
the local quantity
\beq
        R_n =  \sum_{\m} \mid \mbox{\Tr}
               \left\{ G(n)U_\m(n) \right\} \mid^2
          + \sum_{\m} \mid \mbox{\Tr} \left\{ U_\m(n-\hat{\m})G^\dg(n) \right\} \mid^2 
        \hspace{0.5 cm} .
\eeq
Each gauge transformation $G$ is obtained from a SU(2) matrix $g=g_4\Id - \imath g_i \sigma_i$ 
which is then included in one of the $N(N-1)/2$ SU(2) subgroups of SU(N). 
Doing the update in this way $R_n$ can be written as,
\bea
        R_n =  \sum_{i,j=1}^4 \oh g_i a_{ij} g_j
                                                   - \sum_{i=1}^4 g_i b_i + c 
                                  \hspace{0.5 cm}  , 
\eea
being $a_{ij}$ a $4 \times 4$ real symmetric matrix, $b_i$ a real 4-vector
and $c$ a real constant, all depending on the links variables $U_\m(n)$ and 
$U_\m(n-\hat{\m})$. We have to find the maximum of this quantity with the 
constraint that $g$ is a SU(2) matrix. This can be done using standard numerical
algorithms.

Once we obtain the SU(2) matrix $g$ (and then the SU(N) matrix $G$)
maximizing $R_n$, we update the $U_{\mu}$ variables touching the site $n$.
We repeat this procedure over the $N(N-1)/2$ SU(2) subgroups of SU(N) and
over all lattice points. When the whole lattice is covered once we say
we have performed one center gauge fixing sweep. We make a number of
center gauge fixing sweeps on a lattice configuration and stop the
procedure when the quantity $R$ is stable within a given precision.

\section{Fixing SU(3) vortex-like configurations.}   

In this section we want to check the method of center gauge fixing
presented in the previous section. To this purpose we apply the method
to previously prepared SU(3) vortex-like configurations.
Finally, we study the structure of the gauge fixed solution after center
projection and also compare with the results obtained for the SU(2) case.

We use the methods of previous work on vortex-like solutions
in SU(2) Yang-Mills theory \cite{R2xT2} to obtain similar configurations
for the SU(3) group. The main result in \cite{R2xT2} is that there are solutions of
the SU(2) Yang-Mills classical equations of motion in $R^4$ which are self-dual
and vortex-like. These solutions are constructed from self-dual $R^2 \times T^2$ 
configurations by glueing to themselves the two periodic directions.
The method to obtain these  $R^2 \times T^2$ configurations is the one described in references
\cite{cool1,R2xT2}. These solutions are obtained using the lattice aproach
and a standard cooling algorithm. For the SU(2) case we have solutions obtained on lattices of sizes
$N_l \times N_s \times N_l \times N_s$ ($t,x,y,z$ directions, respectively)
with $N_s=4,5,6,7$ and $N_l=4 \times N_s$, and using twisted boundary conditions given 
by the vectors $\vec{k} = \vec{m} = (0,1,0)$. Their properties are described in \cite{R2xT2}.
For the SU(3) case we work on lattices of sizes
$N_l \times N_l \times N_s \times N_s$ with $N_s=4,5,6$ and $N_l=6 \times N_s$ and 
impose twisted boundary conditions given by the vectors $\vec{k} = \vec{m} = (1,0,0)$. 
The resulting SU(3) configurations are, as expected, localized in the two large directions, t and x, and
almost flat in the two small directions, y and z.
To show this property we calculate the energy profiles $\Epsilon_{\mu}(x_{\mu})$,
defined as the integral of the action density in three coordinates.
We observe that $\Epsilon_{\mu}(x_{\mu})$, for $x_{\mu}=t,x$ is localized in a region of length about 3 (we fix 
the length in the small directions equal to 1) with an exponential fall off at the tails.
In the other two directions: $(y,z)$,
$\Epsilon_{\mu}(x_{\mu})$ is almost flat with only one maximum and one minimum
and varying less than the $10\%$ from the mean value of this function to the
value of the maximum or the minimum.

The other quantity characterizing a vortex is the Wilson loop $W_C(r)$ around this object,
being the path $C$ an $r \times r$ square loop in the xt plane centered at the maximum of the solution. 
We parametrize $W_C(r)$ by the functions $L(r)$ (its module) and $\phi(r)$ (its phase). In figure 1 we plot 
$L(r)$ and in figure 2 $\phi(r)$, both as a function of r. We also plot in the same
figures the values of $L(r)$ and $\phi(r)$ when the path $C$ is a square loop of size $r \times r$
centered at the minimum of the solution in the two small directions, y and z, and at the
maximum in the other two. First, we can see that these functions are almost independent
of coordinates y and z, and second, when $r$ is bigger than the size of the object we obtain
$W_C(r) \sim exp(i 4\pi /3) $, an element of the group center. These are the two expected
properties for a vortex.

Once we have isolated the solutions we apply to them the method of center gauge fixing.
We perform center gauge fixing sweeps over the lattice configurations and stop when the
quantity $R$ is stable up to the 8th decimal figure (after $5000$ sweeps this property is
allways satisfied) . For the SU(3) vortex-like solutions we have worked with we have obtained
the following maximum values: $R=0.90123970,0.93246431$ and $R=0.94918155$,
for the lattices sizes: $N_s=4,5$ and $N_s=6$, respectively.
For the configuration with size $N_s=5$ we have repeated four times the procedure
after applying random gauge transformations over the gauge fixed links and obtained 
new values of $R$: $R=0.93084223,0.93170763,0.93236694,0.93246431$. Note that these
values correspond to different Gribov copies of the solution.
Now we explain how the solutions look like when center-projected.

We project the $SU(N)$ link variables to $Z_N$ link variables using
the standard procedure. From these $Z_N$ variables we calculate the values
of the plaquettes in the xt plane (to compare with the Wilson loop presented
before). For the gauge fixed configurations with maximum value of $R$ we always
obtain a plaquette structure in the xt plane (for all y,z points) with
only two plaquettes different from  the identity. The first one takes the value $exp\{i2\pi /3\}$ and
is located at the top-right corner. This reflects the use of twisted boundary conditions: this plaquette
is multiplied by $Z_{\mu \nu}^* = exp\{i2\pi n_{\mu \nu}/3\}$ to this end.
Then, as it is located on a place with very
small action (far from the core of the solution) $Z_{\mu \nu}^{*} U_{\Box} \sim \Id $
$\rightarrow$ $U_{\Box} \sim Z_{\mu \nu}\Id = Z_{14}\Id = exp\{i2\pi /3\}\Id$
the value obtained after center projection. The other
one is located near the maximum in the action density of the 
solution and has the same value of the Wilson
loop shown in figures 2,3 when the size of the loop is much bigger than the size
of the solution. Now we compare the position of this second plaquette for the center-projected
configurations with maximum values of R and the position of the maximum of the solution.

In table 1 we give the position of the maximum in the action density (Max. Pos.) and
the position of the second plaquette (Pla. Pos.) for the three SU(3) configurations constructed,
identified by their sizes, ($N_l$,$N_s$). We observe that the position of the non trivial plaquette
is very near to the position of the maximum. For the SU(2) case the results obtained are very similar 
(these results are obtained using the method of direct center gauge fixing described in reference \cite{green3}).
In the yt plane, the plane with two large directions, the structure is the same for all
xz points with only two plaquettes different form the identity: one related with the  
use of twisted boundary conditions and the other located near the maximum in the action
density, both taking the value $-1$.
Also in table 2 is given the position of the maximum in the action density and
the position of the second plaquette for the SU(2) configurations studied.

To understand more clearly what is happening we study the four Gribov copies
obtained for the SU(3) configuration with $(N_l,N_s)=(30,5)$. 
The Gribov copies of this configuration and bigger values of $R$,
$R=0.93170763,0.93236694$ and $R=0.93246431$ have the same structure in the xt plane 
(the one described before) with the second plaquette at positions: $(n_x,n_t) = (22,4),
(23,3)$ and $(24,2)$ respectively. We do not find the expected result. We expect that 
the one with biggest $R$ has the same position for the second plaquette and
the maximum of the solution (position $(n_x,n_t) = (22,4)$ ). Actually, the one with the
smallest value of $R$ has the same position for both objects. We understand this by calculating
the quantity $R$ only for links belonging to the t and x directions ($R_{tx}$), obtaining:
$R_{tx}=0.99833177, 0.99806162$ and $R_{tx}=0.99765530$. As expected,
for bigger value of $R_{tx}$ the position of the second plaquette is nearer
to the maximum of the solution.

We also address the structure of the Gribov copy with the smallest value of $R$ ($R=0.93084223$).
In this case the plaquette structure of the xt plane is different for different values of
y,z.  There are twenty-five xt planes. Twenty-one of them have the same structure as before with
the second plaquette located at position $(n_x,n_t) = (6,17)$. Three of them have also the same structure but
at different positions: two at $(n_x,n_t) = (6,16)$ and one at $(n_x,n_t) = (5,17)$.
And the last one has five non trivial plaquettes, the twist carrying plaquette, 
three with value $exp(i 2\pi /3)$ (positions $(n_x,n_t)=(6,17),
(8,30),(25,12)$) and one with value $exp(i 4\pi /3)$ (position $(n_x,n_t)=(27,29)$). In this case there
is no clear relation between the underlying structure of the SU(3) solution and the
structure of the center-projected one.

Finally, to clarify the role of the twist carrying plaquette, we study a zero action
configuration for the SU(2) group satisfying twisted boundary conditions given by
the twist vectors $\vec{k} = \vec{0}$ and $\vec{m}=(0,1,0)$. In this case the 
continuum solution is known: $A_{\mu}=0$ and $\Gamma_1=\imath \sigma_1$, $\Gamma_3=\imath \sigma_3$,
$\Gamma_2=\Gamma_4=\Id$ (being $A_{\mu}$ the gauge field and $\Gamma_{\mu}$ the twist matrices).
It is also known the lattice version of this solution, all links equal to the identity except
the following ones: $U_1(n_x=N_x)= \imath \sigma_1$, $U_3(n_z=N_z)= \imath \sigma_3$. 
We built this configuration on a lattice ${N_s}^4$ with $N_s=6$ and
apply the center gauge fixing procedure to it, obtaining that it is 
a local maximum of the functional $R$ with value $R=0.91666666$. Note that the center projected
configuration has all plaquettes equal to the identity. To obtain other value of R
we apply first a random gauge transformation and then the center gauge fixing procedure.
We can obtain with this method different copies of the solution with the same value
of $R$, $R=0.95603917$. The center projected configuration has the following properties:
all plaquettes equal to 1 in the planes xy, xt, yz, yt and zt; in the xz plane
all plaquettes equal to 1 except two, one the twist carrying plaquette at the top-right
corner of the plane, and the other in any place, both taking the value -1.   
We have repeated this procedure six times and the second plaquette appears at points:
$(n_x,n_z)=(5,6);(6,5);(6,4);(1,3);(2,3);(2,1)$. This result reflects that there is
no special point in the underlying solution.
 
We can conclude saying that twisted boundary conditions forces to  have one 
non trivial plaquette at the position of the twist carrying plaquette (could be any 
position in the plane).  But you can not
have only one non trivial plaquette in the center projected configuration, 
you must have al least two. Then the other non trivial 
plaquette, if there are only two, is located in other place of the lattice in the vecinity
of points with higher action density: at any place for the zero action solution and near the
maximum in the action density for the fractional charge solutions. The result for the
zero action solution implies that not allways a P-vortex is associated with the core of
a center vortex.

\section{Conclusions}   

We have presented a new method of center gauge fixing to maximal center gauge
and checked that it works well. We have used this method to study the structure 
of SU(2) and SU(3) vortex-like solutions in the maximal center gauge when center-projected.
It seems that, for the fractional charge solutions with maximum value of $R$, there is a clear
relation between the structure of the SU(2) and SU(3) configurations and the structure of the 
center-projected ones. The Wilson loop characterizing the vortex-like solution, calculated
in the plane in which it is localized, takes the value of an element of the group center
when the size of the loop is much bigger than the size of the solution. After center
projection, this plane has only two plaquettes different from the identity, one forced
by twisted boundary conditions and the other taking
the same value of the element of the group center characterizing the vortex solution.
This second plaquette is localized near the maximum in the action density of the solution.
We can say that these objects share the same structure as the ones described in
\cite{green1,green2,green3,lang,lang2,cher1,cher2,forcrand}, and therefore are good candidates
to be confining configurations.

There are a few remarks regarding our results. 
First, with our method we do not know if there are other
Gribov copies with bigger values of $R$, we only know that the configurations
obtained are a local maximum of equation (\ref{mesonlike}). Second, the picture
obtained for the copies with maximum value of $R$ seems to be independent of 
the lattice spacing, but, in our opinion, this property will be changed for smaller
lattice spacing. We think that, for smaller lattice spacing, the size of the
vortex solution will become relevant and then a bigger number of plaquettes in
the neighbourhood of the core of the solution will be different from the identity
(nevertheless, the appropriate value for the Wilson loop characterizing the
vortex will still be obtained with a sufficiently large loop size).
Third, we have seen that there exist other Gribov copies (with
smaller value of $R$) in which the relation between the underlying SU(N) structure
and the projected one is not clear. This means that it is quite important to obtain
a value of $R$ big enough to see a relation between the projected and the unprojected
configuration, and a more systematic way to identify the vortex properties, working 
independently of previously known features, will be required. The recent work in 
reference \cite{forcrand2} could be the way to solve these problems.

\section*{Acknowledgements}
I  acknowledge useful conversations with M. Garc\'{\i}a P\'erez, 
A. Gonz\'alez-Arroyo and C. Pena. The present work was financed 
by CICYT under grant AEN97-1678.

\newpage

\begin{table}
\begin{center}
\caption{ {\footnotesize Position of the maximum in the accion density (Max. Pos.) and
position of the non trivial plaquette (Pla. Pos.) for the SU(N) configurations studied.}}
\vspace{0.1 cm}
\begin{tabular}{||c||c||c||c||}
\hline
{\bf N} & {\bf Size} ($N_l$,$N_s$) &  {\bf Max. Pos.} ($n_y$,$n_t$)  &  {\bf Pla. Pos.} ($n_y$,$n_t$) \\  \hline 
2 &(16,4)                  &   (11,7)                        &  (11,7)                        \\  \hline 
2 &(20,5)                  &   (1,9)                         &  (20,8)                        \\  \hline 
2 &(24,6)                  &   (2,11)                        &  (1,11)                        \\  \hline 
2 &(28,7)                  &   (5,16)                        &  (6,14)                        \\  \hline \hline

{\bf N} & {\bf Size} ($N_l$,$N_s$) &  {\bf Max. Pos.} ($n_x$,$n_t$)  &  {\bf Pla. Pos.} ($n_x$,$n_t$)  \\ \hline 
3 &(24,4)                  &   (21,16)                       &  (22,16)                        \\  \hline 
3 &(30,5)                  &   (22,4)                        &  (24,2)                         \\  \hline 
3 &(36,6)                  &   (14,35)                       &  (17,34)                        \\  \hline \hline
\end{tabular}
\end{center}
\end{table}

\newpage

\begin{figure}
 \caption{ The module $L(r)$ of the Wilson loop $W_C(r)$ 
            is shown as a function of $r$ for the solutions with lattice sizes $N_s=4,5,6$.
            Full symbols correspond to Wilson loops centered at the maximum of the solutions 
            and empty symbols to the same quantity centered at the minimum of the solutions
            in the $y,z$ directions and at the maximum in the $t,x$ directions.}
 \vbox{ \vskip -1.5 cm \hskip -0.3 cm \hbox{  \epsfxsize=450pt \hbox{\epsffile{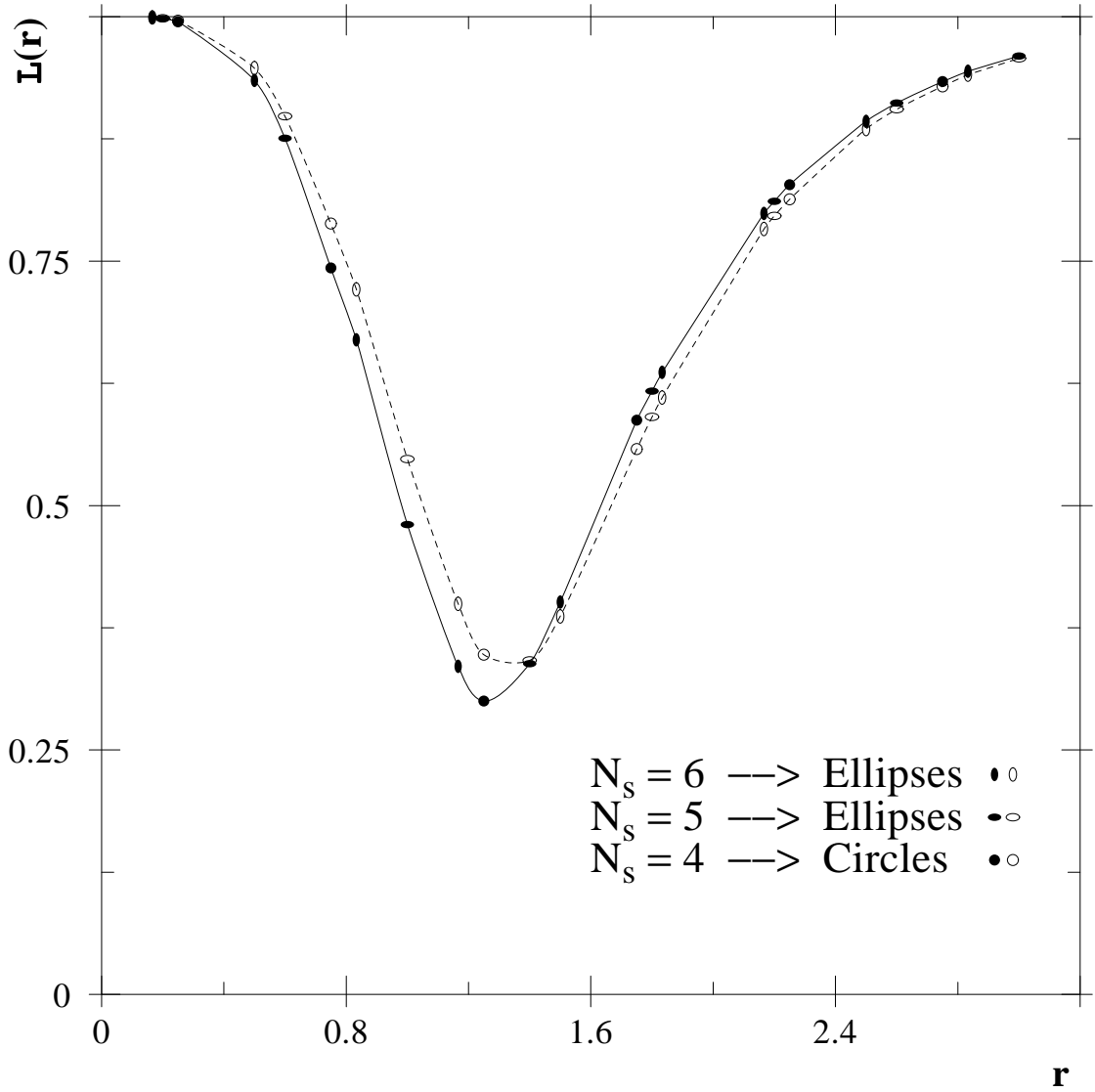} } } } 
\end{figure}

\newpage

\begin{figure}
 \caption{ The phase $\phi(r)$ of the Wilson loop $W_C(r)$  
           is shown as a function of $r$ for the solutions with lattice sizes $N_s=4,5,6$.
            The symbols have the same meaning as in figure 1.}
 \vbox{ \vskip -1.5 cm \hskip -0.3 cm \hbox{  \epsfxsize=450pt \hbox{\epsffile{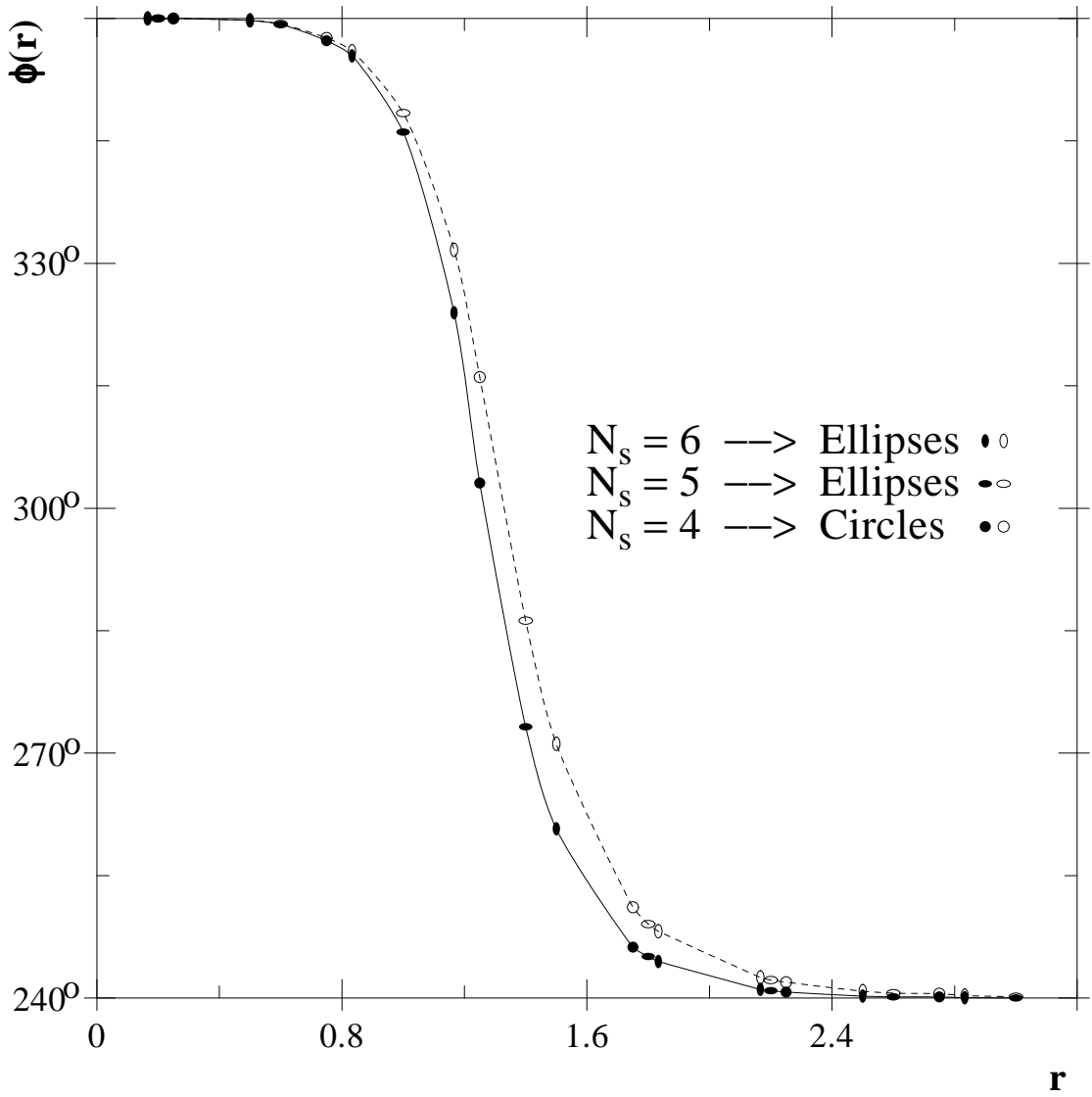} } } } 
\end{figure}

\end{document}